\pdfoutput=1
\documentclass[twoside,slac_one]{revtex4}
\usepackage{graphicx}
\usepackage{fancyhdr}
\usepackage{amsmath} 
\usepackage{bm}
\usepackage{amsxtra}
\usepackage{amssymb}
\usepackage{amsthm}
\usepackage{latexsym}
\usepackage{lscape}

\begin{document}

\title{The Wake of a Quark Moving through 
Hot QCD Plasmas vs. $\mathcal{N}=4$ SYM Plasmas}

%

\author{Juhee Hong}
\affiliation{Department of Physics and Astronomy, Stony Brook University, Stony Brook, NY, USA}

\begin{abstract}
We present the energy density and flux distribution of a heavy quark 
moving through high temperature QCD plasmas and 
compare them with those in the strongly coupled $\mathcal{N}=4$ SYM plasma.

The Boltzmann equation is reformulated as a Fokker-Planck equation 
at the leading log approximation and is solved numerically with 
nontrivial boundary conditions in momentum space. 
We use kinetic theory and perform a Fourier transform to calculate 
the energy and momentum density in position space. 
The angular distributions exhibit the transition to the ideal hydrodynamics 
and are analyzed with the first and second order hydrodynamic source.

The AdS/CFT correspondence allows the same calculation at strong coupling. 
Compared to the kinetic theory results, the energy-momentum tensor is better described 
by hydrodynamics even after accounting for the differences in the 
shear viscosities. 
We argue that the difference between the Boltzmann equation and 
the AdS/CFT correspondence comes from 
the second order hydrodynamic coefficient $\tau_\pi$ which is generically 
large compared to the shear length in a theory based on the Boltzmann equation. 
\end{abstract}

\maketitle

\thispagestyle{fancy}


\section{Introduction}

When a heavy quark moves supersonically through plasmas, 
the energy and momentum are redistributed, 
producing a Mach cone and diffusion wake. 
Originally, the study of the Mach cone structure was motivated 
by an attempt to understand the 
two particle correlation functions measured in the heavy ion collisions. 
Although the Mach cone failed to explain the experimental observation, 
it is a simple way to study the medium response to an energetic particle.   

Using the AdS/CFT correspondence, the stress-energy tensor 
in the strongly coupled SYM plasma due to a heavy quark have been 
calculated \cite{Gubser:2007ga, Chesler:2007sv}.
In this work, we study the Mach cone produced in the high temperature  
QCD plasmas at 
weak coupling and compare it with the strong coupling results 
given by the AdS/CFT correspondence.  

\section{Linearized Boltzmann Equation at Leading Log}
We start with the Boltzmann equation  
\begin{equation}
\left(\partial_t+\mathbf{v}_p\cdot\partial_{\mathbf{x}}\right) f(t,\mathbf{x},\mathbf{p})
=C[f,\mathbf{p}] \, , 
\end{equation}
where the particle distribution function is linearized around the equilibrium, 
$f(t,\mathbf{x},\mathbf{p})=n_p+\delta f(t,\mathbf{x},\mathbf{p})$ 
with $n_p=1/(e^{p/T}-1)$. 
The collision term is given by 
\begin{equation}
C[f,\mathbf{p}]=-\frac{1}{2}\int_{\mathbf{k},\mathbf{p}',\mathbf{k}'}
\arrowvert M \arrowvert ^2 (2\pi)^4\delta^4(P+K-P'-K')
\bigg[f(\mathbf{p})f(\mathbf{k})[1+f(\mathbf{p}')][1+f(\mathbf{k}')]
-f(\mathbf{p}')f(\mathbf{k}')[1+f(\mathbf{p})][1+f(\mathbf{k})]\bigg] \, , 
\end{equation}
where momentum space integrals are abbreviated as $\int_{\mathbf{p}}\equiv \int \rm{d}^3\mathbf{p}/(2\pi)^3$. 
At the leading log approximation, we consider only t-channel $2\rightarrow 2$ scattering. 
Two hard particles with momenta of order $T$  
are scattered into two others, exchanging soft particles with momenta 
of order $gT$, where $g$ is the weak coupling constant. 
The particles are scattered with angles of order $g$. 
With the definition $\delta f(\mathbf{p}) \equiv n_p(1+n_p)\chi(\mathbf{p})$, 
the collision integral can be simplified and expressed as a variational form 
\footnote{The dimension of the adjoint representation is $d_A=N_c^2-1$. 
The Casimirs of the adjoint and fundamental representations are 
$C_A=N_c$ and $C_F=(N_c^2-1)/2N_c$.  
The Debye mass is $m_D^2=2g^2C_A\int_{\mathbf{p}}
\frac{n_p(1+n_p)}{T}=\frac{g^2T^2}{3}N_c$.}
\cite{Arnold:2000dr, Hong:2010at}
\begin{eqnarray}
\left(\partial_t+\mathbf{v}_p\cdot\partial_{\mathbf{x}}\right) \delta f(t,\mathbf{x},\mathbf{p})
&=&T\mu_A \frac{\partial}{\partial p^i}\left(  n_p(1 + n_p) 
\frac{\partial} {\partial p^i} 
\left[ \frac{\delta f }{n_p (1 + n_p)} \right] \right) 
+ \mbox{gain terms} \, , \nonumber\\
\mu_A &\equiv & \frac{g^2C_A m_D^2 }{8 \pi} \log\left(\frac{T}{m_D}\right).
\end{eqnarray}
Without gain terms, the equation can be reformulated as 
a Fokker-Planck equation
\begin{equation}
 (\partial_t + {\mathbf{v}_p}\cdot \partial_\mathbf{x}) \delta f(t,\mathbf{x},\mathbf{p}) = -\mu_A \,(1+ 2n_p) \, \hat{\mathbf{p}} \cdot \frac{\partial \delta f }{\partial \mathbf{p}} \, + \,  T\mu_A \nabla_\mathbf{p}^2 \,\delta f 
\end{equation}
and the motion of the particle excess, $\delta f$, can be described 
by the Langevin equation with the drag and random  kicks ${\bm \xi}(t)$ 
\begin{equation}
 \frac{dp^i}{dt} = - \mu_A (1 + 2n_p ) \hat p^i  + \xi^i(t) \, ,   \qquad \langle \xi^i(t) \xi^j(t') \rangle = 2 T\mu_A \delta^{ij} \delta(t-t') \, .
\end{equation}
In this diffusion process, the excess particles lose their energy and momentum to the plasma. 
However, there are work and momentum transfer on the particles per time, per degree of freedom, per volume
\begin{eqnarray}
 \frac{\rm{d} E}{\rm{d} t} &\equiv&   -T\mu_A \int  \frac{\rm{d}^3\mathbf{p}}{(2\pi)^3 }
 \, n_p(1+n_p) \hat{\mathbf{p}} \cdot
   \frac{\partial \chi}{\partial \mathbf{p} }  \, , \nonumber\\
 \frac{\rm{d} P}{\rm{d} t} &\equiv&   -T\mu_A \int  \frac{\rm{d}^3\mathbf{p}}{(2\pi)^3 }
 \, n_p(1+n_p)
   \frac{\partial \chi}{\partial \mathbf{p} }  \, .
\end{eqnarray}
In terms of these energy and momentum transfer, the gain terms are given by
\begin{equation}
\mbox{gain terms} =  
  \frac{1}{\xi_B} 
\left[ \frac{1}{p^2}\frac{\partial}{\partial p}p^2 n_p(1 + n_p)\right] \frac{\rm{d} E}{\rm{d} t} 
  + \frac{1}{\xi_B} \left[\frac{\partial}{\partial \mathbf{p}} n_p (1 + n_p)  \right]  \cdot \frac{ \rm{d}{\bm P}}{\rm{d} t}  \, ,
\end{equation}
where we defined 
$\xi_B \equiv \int \frac{\rm{d}^3 \mathbf{p}}{(2\pi)^3 } n_p(1 + n_p) 
= \frac{T^3}{6}$.
Using the Boltzmann equation with the gain terms, 
it is easy to show that the lost energy and momentum of the excess particles 
are exactly compensated by the gain terms. 
Therefore, the total energy and momentum are conserved during the evolution.

In order to get the boundary condition, we consider the excess of soft gluons within a small ball of radius $\Delta p \sim gT$ centered at $\mathbf{p}=0$
\begin{equation}
 \int_{p=0}^{p=\Delta p} \frac{\rm{d}^3\mathbf{p}}{(2\pi)^3} 
 n_p(1+n_p)\chi(\mathbf{p})   \simeq \frac{T^2}{2\pi^2} 
\chi({\bm 0}) \Delta p \, 
\end{equation}
which is vanishing at the weak coupling limit. 
Since it is easy to emit or absorb a soft gluon, we choose the absorptive boundary condition \cite{Arnold:2006fz}
\begin{equation}
  \left. \chi(\mathbf{p})  \right|_{\mathbf{p}\rightarrow 0 } = 0 \, .
\end{equation}
As a result, there is flux ($\partial \chi/\partial\mathbf{p}$)
 at $\mathbf{p}=0$ and the particle number changes.

Using the linearized Boltzmann equation with the absorptive boundary condition, 
we determined the transport coefficients including the shear viscosity $\eta$ 
and the second order hydrodynamic coefficient $\tau_\pi$ \cite{Hong:2010at}
\begin{equation}
\frac{\eta}{sT}= 0.4613\frac{T}{\mu_A} \, \, , \qquad
\tau_\pi = 6.32 \frac{\eta}{sT} \,  \,  \qquad \qquad \mbox{(Boltzmann)} \, ,
\end{equation}
which will be used in the Mach cone analysis in the following. 
In comparison, we provide the corresponding coefficients for 
the AdS/CFT correspondence \cite{Kovtun:2004de, Baier:2007ix}
\begin{equation}
\frac{\eta}{sT}= \frac{1}{4\pi T} \, \, , \qquad
\tau_\pi = \frac{2-\ln 2}{2\pi T} \,  \,  \qquad \qquad \mbox{(AdS/CFT)} \, .
\end{equation}

\section{Mach Cone}
We simulate a Mach cone by solving the linearized Boltzmann equation with a source term 
\begin{equation}
\left(\partial_t+\mathbf{v}_p\cdot\partial_{\mathbf{x}}\right) \delta f(t,\mathbf{x},\mathbf{p})
=C[f,\mathbf{p}] +S(t,\mathbf{x},\mathbf{p}) \, .
\end{equation}
In the presence of a heavy quark moving with a constant velocity $\mathbf{v}$, 
the particles in equilibrium are scattered, producing the source around the quark, $S(t,\mathbf{x},\mathbf{p})=S(\mathbf{p})\delta^3(\mathbf{x}-\mathbf{v} t)$. 
The source term is given by another collision integral
\begin{equation}
S(\mathbf{p})=
-\frac{1}{2}\int_{\mathbf{k},\mathbf{p}',\mathbf{k}'}\arrowvert M \arrowvert ^2
(2\pi)^7\delta^4(P+K-P'-K')
\bigg[f(\mathbf{p})[1+f(\mathbf{p}')]\delta^3(\mathbf{k}-\mathbf{k}_H)
-f(\mathbf{p}')[1+f(\mathbf{p})]
\delta^3(\mathbf{k}'-\mathbf{k}_H)\bigg] \, ,
\end{equation}
where we used 
$f_H(\mathbf{k})=(2\pi)^3\delta^3(\mathbf{k}-\mathbf{k}_H)\delta^3(\mathbf{x}
-\mathbf{v} t)$ as the distribution function of the heavy quark. 
Especially, when the quark moves with the speed of light, $v=1$, 
it is simplified at the leading log order 
\begin{eqnarray}
S(\mathbf{p})&=&\frac{\mu_F}{2d_A\xi_B}n_p(1+n_p)
\left[\left(-\frac{2}{p}+\frac{1+2n_p}{T}\right)+\frac{(1+2n_p)}{T}\hat{\mathbf{p}}\cdot\mathbf{v}\right] \, ,\nonumber\\
\mu_F &\equiv & \frac{g^2C_F m_D^2 }{8 \pi} \log\left(\frac{T}{m_D}\right) \, .
\end{eqnarray}

\begin{figure}[h]
\centering
\includegraphics[width=80mm]{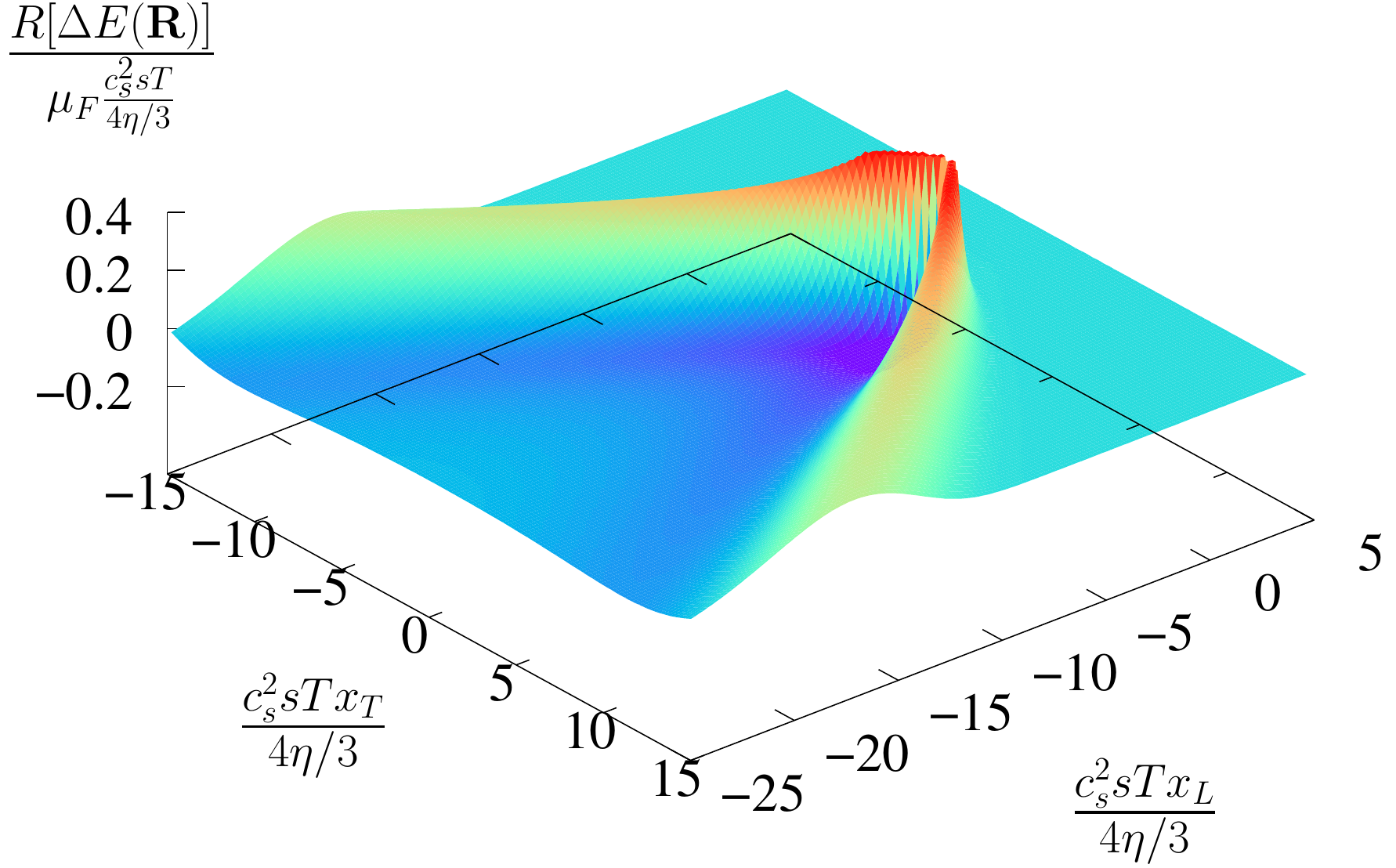}
\includegraphics[width=80mm]{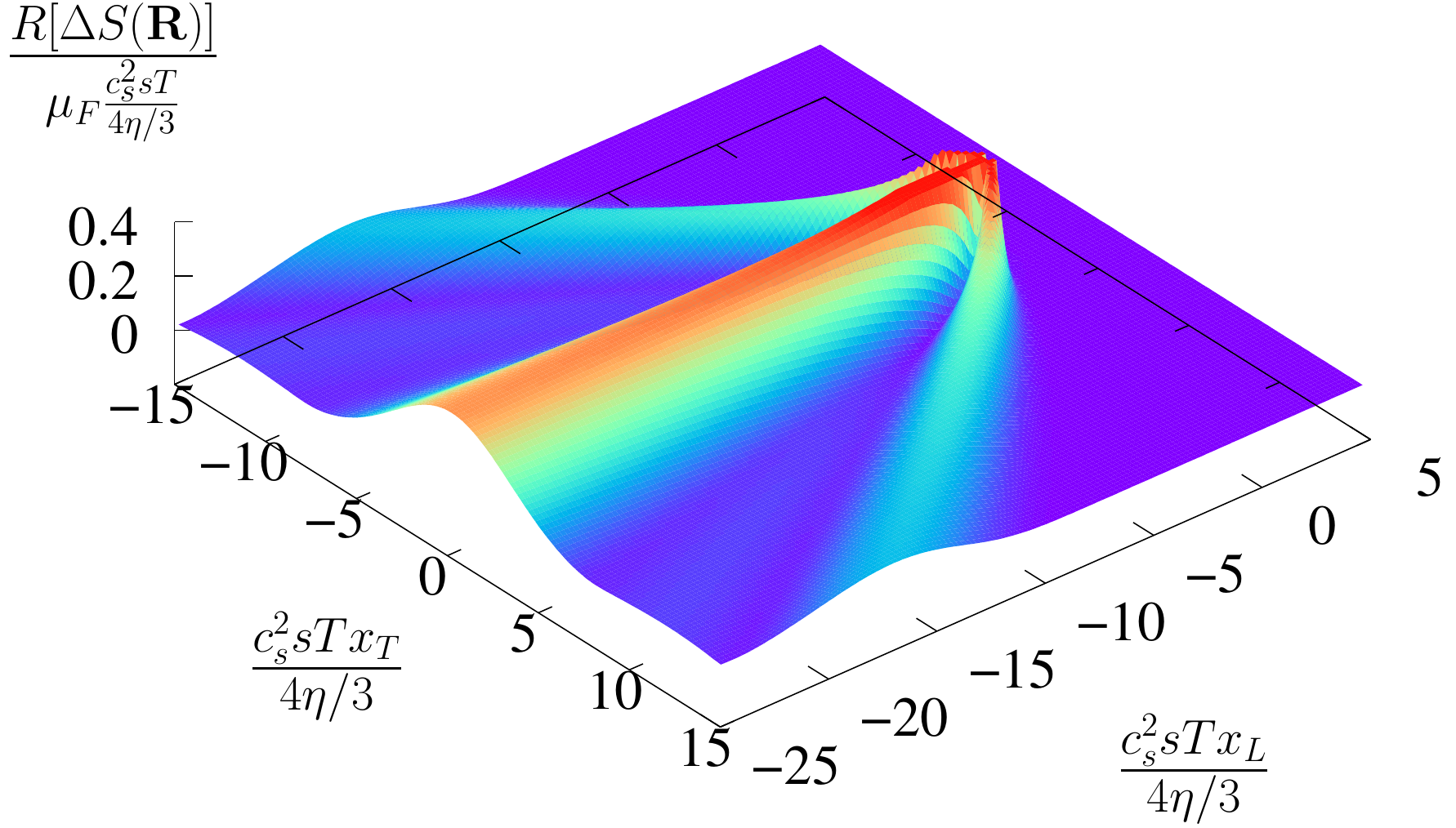}
\caption{(a) The energy density distribution,  
$R \, [\Delta E(\mathbf{R})]/[\mu_F(c_s^2sT)/(4\eta/3)]$
 and (b) the energy flux distribution, 
$R \, [\Delta S(\mathbf{R})]/[\mu_F(c_s^2sT)/(4\eta/3)]$
given by the Boltzmann equation. 
They show a smooth transition from the free streaming to 
the hydrodynamic regime. 
We can see a Mach cone and a diffusion wake (in the flux distribution) behind the heavy quark. 
($\mathbf{R}=x_T\hat{\mathbf{x}}+x_L\hat{\mathbf{z}}$)} 
\label{boltzfull}
\end{figure}

\begin{figure}[h]
\centering
\includegraphics[width=80mm]{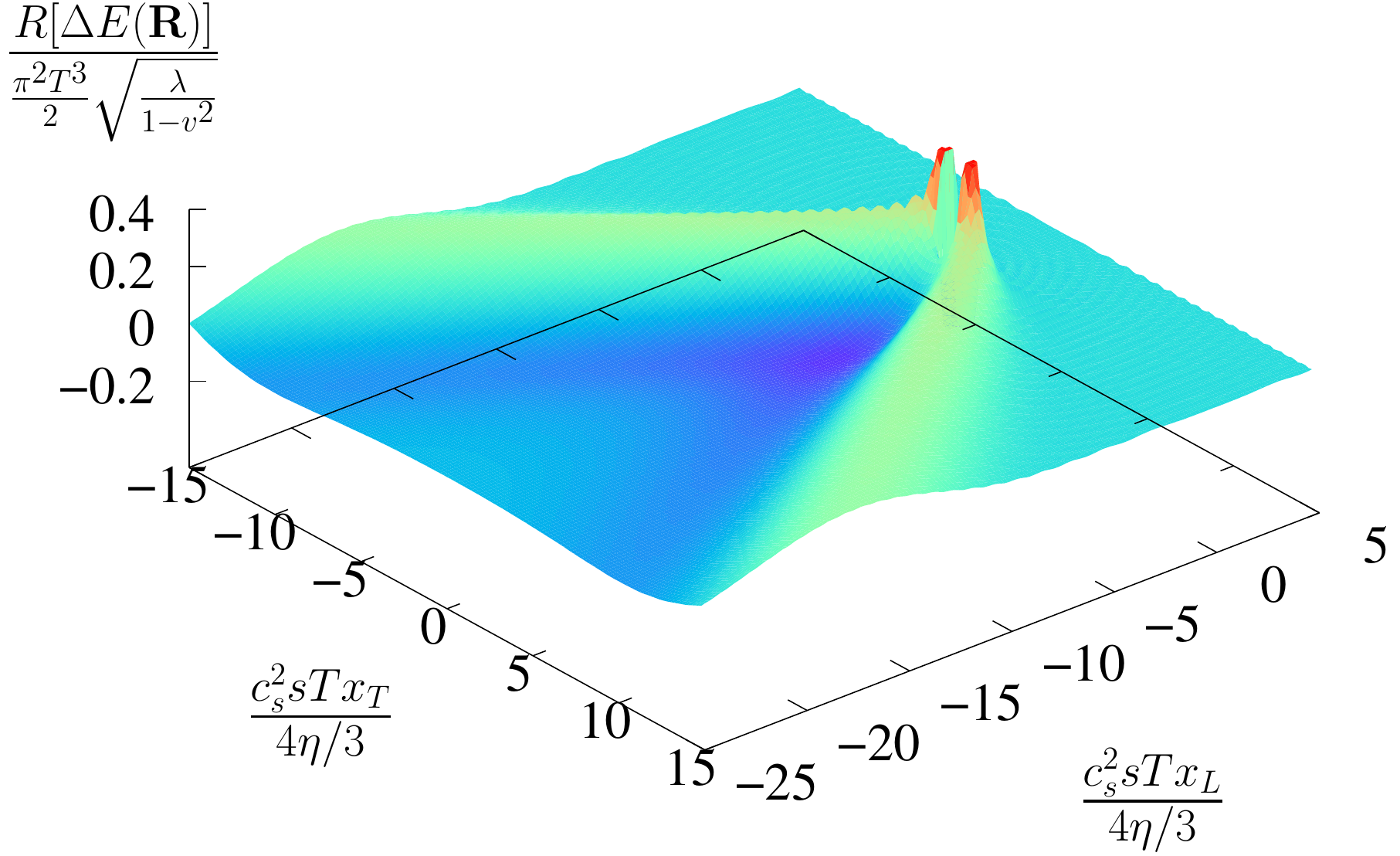}
\includegraphics[width=80mm]{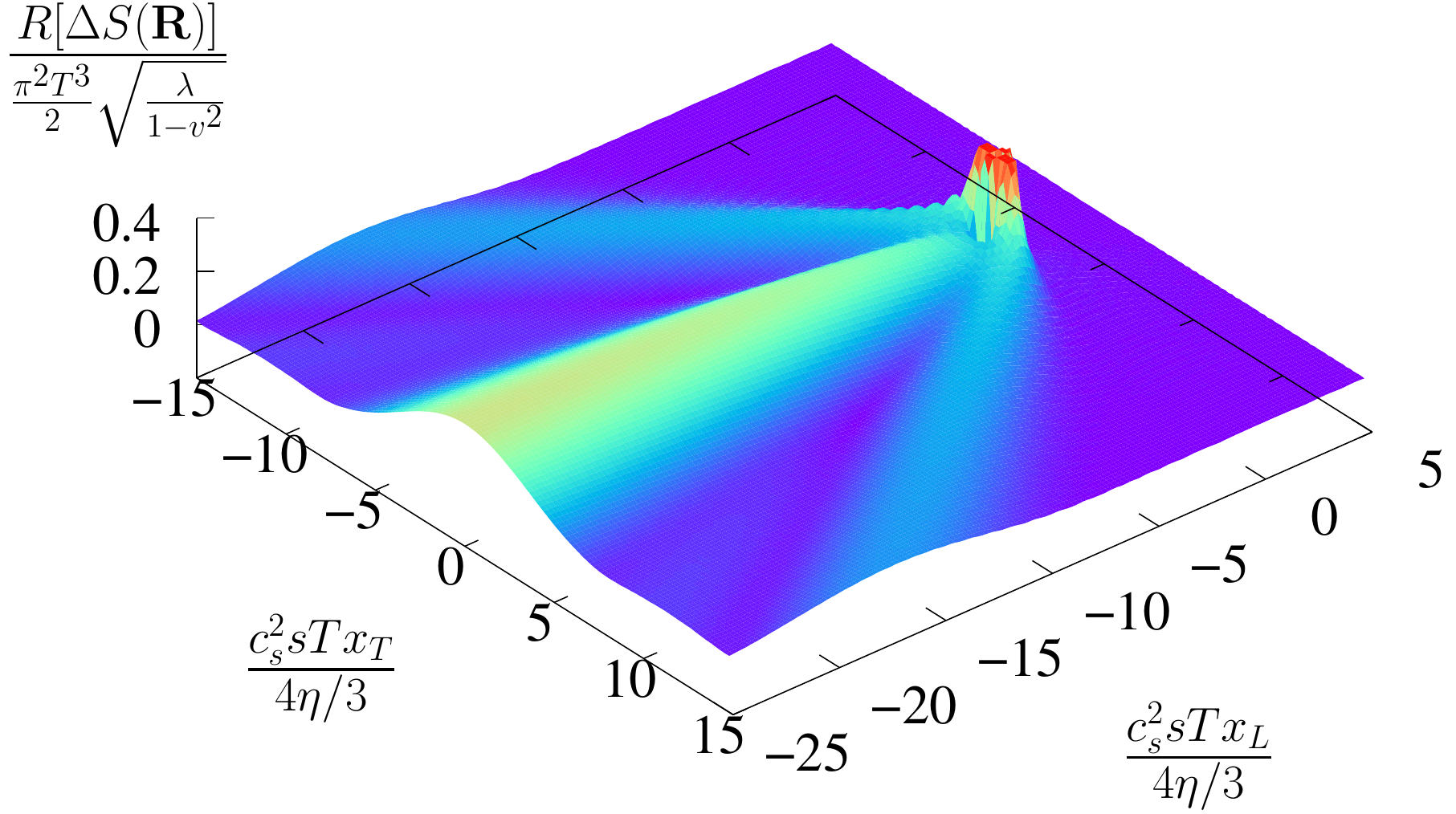}
\caption{(a) The energy density distribution, 
$R[\Delta E(\mathbf{R})]/[\pi^2 T^3 \sqrt{\lambda /(1-v^2)}/2]$
and (b) the energy flux distribution, 
$R[\Delta S(\mathbf{R})]/[\pi^2 T^3 \sqrt{\lambda /(1-v^2)}/2]$
given by the AdS/CFT correspondence. 
They are singular near the heavy quark at origin.
We can see a Mach cone and a diffusion wake (in the flux distribution) behind the heavy quark. 
($\mathbf{R}=x_T\hat{\mathbf{x}}+x_L\hat{\mathbf{z}}$)} 
\label{adsfull}
\end{figure}

Our strategy to get a Mach cone is the following: 
(i) perform a Fourier transform of the Boltzmann equation and 
solve it in Fourier space, 
(ii) calculate the stress-energy tensor in Fourier space using kinetic theory, 
(iii) perform a Fourier transform of the tensor back to position space to get the distributions. 
We assume the heavy quark moves along $\hat{\mathbf{z}}$ in the lab coordinate system $(x,y,z)$ and 
introduce the Fourier coordinate system $(x',y',z')$, 
where $\hat{\mathbf{z}}'$ points along the Fourier momentum $\mathbf{k}$. 
For simplicity, we use the rotational symmetry around $\hat{\mathbf{z}}$ and 
let $\mathbf{k}$ lie on $zx$ plane. 
Then the source in the Fourier coordinate system becomes 
\begin{eqnarray}
S(\mathbf{p})&=&\sqrt{4\pi}\frac{\mu_F}{2d_A\xi_B}n_p(1+n_p)
\left[\left(-\frac{2}{p}+\frac{1+2n_p}{T}\right)
H_{0,0}(\hat{\mathbf{p}})  +\frac{(1+2n_p)}{\sqrt{3}T}
\bigg(\cos\theta \, H_{1,0}(\hat{\mathbf{p}})
+\sin\theta \, \,H_{1,1}(\hat{\mathbf{p}}) \bigg) \right] \, , \qquad
\label{source}
\end{eqnarray}
where $\theta$ is the angle between $\hat{\mathbf{z}}$ and $\mathbf{k}$  
and $H_{l,m}(\hat{\mathbf{p}})$ 
is the real spherical harmonics defined in Fourier space \cite{Hong:2010at}
\begin{equation}
H_{0,0}(\hat{\mathbf{p}})=\frac{1}{\sqrt{4\pi}} \, \, , \qquad
H_{1,0}(\hat{\mathbf{p}})=\sqrt{\frac{3}{4\pi}} \, 
\hat{\mathbf{p}}^{z'} \, , \qquad
H_{1,1}(\hat{\mathbf{p}})=\sqrt{\frac{3}{4\pi}} \, 
\hat{\mathbf{p}}^{x'} \, .
\end{equation}
The Boltzmann equation in Fourier space is given by 
\begin{equation}
(-i\omega+i\mathbf{v}_p\cdot\mathbf{k})\delta f(\omega,\mathbf{k},\mathbf{p})
=C[\delta f,\mathbf{p}]+2\pi S(\mathbf{p})\delta(\omega-\mathbf{v}\cdot\mathbf{k}) \, . 
\end{equation}
We solve it for 
$\delta f(\omega,\mathbf{k},\mathbf{p}) \equiv 2\pi
\delta(\omega-\mathbf{v}\cdot\mathbf{k})n_p(1+n_p)
\chi(\omega,\mathbf{k},\mathbf{p})$ 
and calculate the stress-energy tensor using kinetic theory
\begin{equation}
\delta T^{0\mu}(\omega,\mathbf{k})=2d_A\int_{\mathbf{p}}p^\mu 
\delta f(\omega,\mathbf{k},\mathbf{p}) \, .
\end{equation}
By Fourier-transforming back to position space, 
we get the energy and momentum density distribution 
\begin{eqnarray}
\delta T^{00}(t,\mathbf{x})&=&2d_A\int_{\omega,\mathbf{k}}
e^{-i\omega t+i\mathbf{k}\cdot\mathbf{x}} \, 
\delta T^{00}(\omega,\mathbf{k}) \, , \nonumber\\
\delta T^{0z}(t,\mathbf{x})&=&2d_A\int_{\omega,\mathbf{k}}
e^{-i\omega t+i\mathbf{k}\cdot\mathbf{x}}
\bigg[\sin\theta \, \delta T^{0x'}(\omega,\mathbf{k})
+\cos\theta \, \delta T^{0z'}(\omega,\mathbf{k})\bigg] \, , \nonumber\\
\delta T^{0x}(t,\mathbf{x})&=&2d_A\int_{\omega,\mathbf{k}}
e^{-i\omega t+i\mathbf{k}\cdot\mathbf{x}}
\bigg[\cos\theta \, \delta T^{0x'}(\omega,\mathbf{k})
-\sin\theta \, \delta T^{0z'}(\omega,\mathbf{k})\bigg] \, , \nonumber\\
\end{eqnarray}
where the integrands inside the square brackets are the real momentum  
$\delta T^{0z}(\omega,\mathbf{k})$ and 
$\delta T^{0x}(\omega,\mathbf{k})$ in terms of Fourier components. 
The numerical results of the energy and the momentum density distribution are shown in ~Figure~\ref{boltzfull}. 
In comparison, we present the AdS/CFT results on the same scale in ~Figure~\ref{adsfull}. 

\begin{figure}[h]
\centering
\includegraphics[width=80mm]{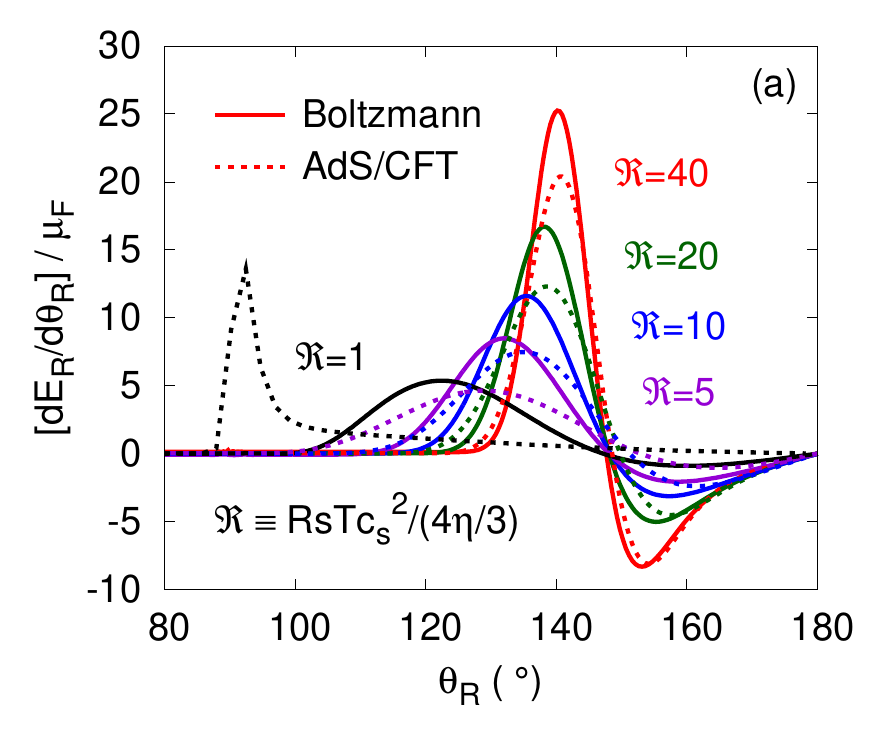}
\includegraphics[width=80mm]{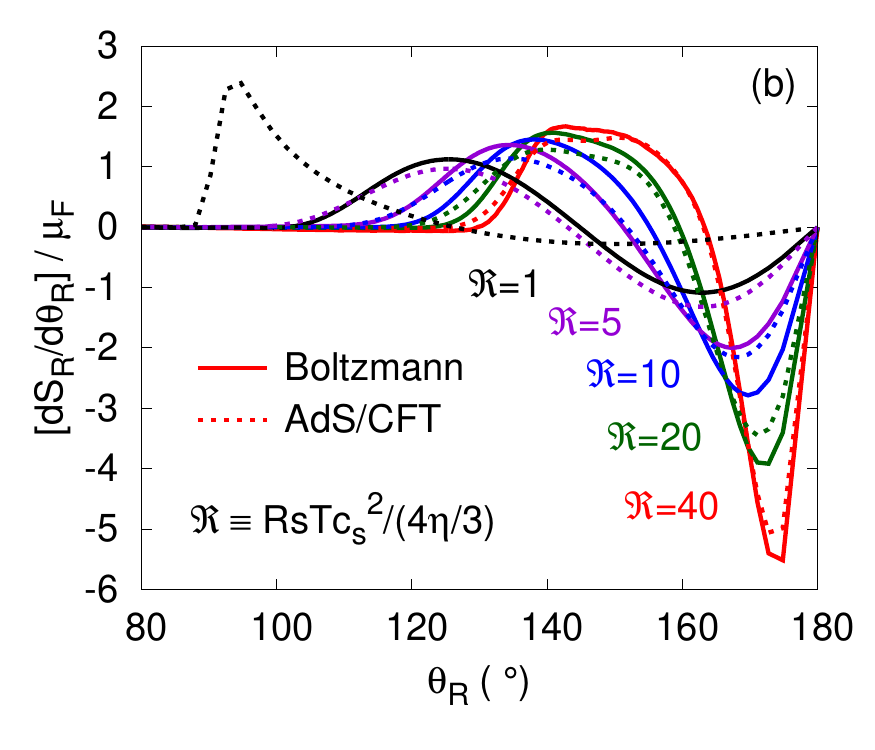}
\caption{The angular distribution of (a) the energy density 
$[\rm{d}E_R/\rm{d}\theta_R]/\mu_F$ and 
(b) the energy flux $[\rm{d}S_R/\rm{d}\theta_R]/\mu_F$ 
at distances $\mathcal{R}=1, \, 5, \, 10, \, 20, \, \mbox{and}\, 40$. 
The Boltzmann results are compared with the AdS/CFT results. 
The distance from the heavy quark 
is measured compared to the shear length, $\eta/(sT)$. 
At $\mathcal{R}=1$, the AdS/CFT results are 
dominated by the zero temperature contribution 
which is absent in the Boltzmann formulation. 
$\mu_F$ for the Boltzmann equation corresponds to 
$\pi T^2\sqrt{\lambda/(1-v^2)}/2$ for the AdS/CFT correspondence.}
\label{comparison}
\end{figure}

In order to analyze the structure, 
we define the angular distribution with 
an angle $\theta_R$ measured relative to $\hat{\mathbf{z}}$ 
on $zx$ plane. 
At a fixed distance 
$R=|x_T \hat{\mathbf{x}} + x_L \hat{\mathbf{z}}|$ from the heavy quark, 
the energy and the momentum density are given by  
\begin{eqnarray}
\frac{\rm{d} E_R}{\rm{d} \theta_R}&\equiv& 2\pi R^2 
\sin\theta_R \, \Delta E(\mathbf{R}) \, , \nonumber\\
\frac{\rm{d} S_R}{\rm{d} \theta_R}
&\equiv & 2\pi R^2\sin\theta_R \, \hat{\mathbf{R}}
\cdot\Delta \mathbf{S}(\mathbf{R}) \, ,
\end{eqnarray}
where $\Delta E \equiv \delta T^{00}$ and $\Delta S_i \equiv \delta T^{0i}$. 
The results are shown in ~Figure~\ref{comparison}, where 
distances ($R$) are measured compared to the shear length, $\eta/(sT)$. 
Generally, the Boltzmann equation and the AdS/CFT correspondence give similar distributions with a Mach cone and a diffusion wake, except for $\mathcal{R}=1$. 
At short distances, the AdS/CFT solutions are dominated by 
the zero temperature contribution which is absent in the Boltzmann formulation.

\section{Hydrodynamics}
Hydrodynamics is an effective theory at long distances and 
is given by the conservation of the energy-momentum tensor  
\begin{equation}
T_{hydro}^{\mu\nu}=[e(T)+\mathcal{P}(T)]u^\mu u^\nu +\mathcal{P}(T)g^{\mu\nu}
+\pi^{\mu\nu} \, ,
\label{Tmunuhydro}
\end{equation}
where the first two terms on the right hand side correspond to 
the ideal hydrodynamics and the last term is the dissipative part. 
The heavy quark perturbs it away from equilibrium 
\begin{equation}
e(t,\mathbf{x}) \simeq e_o +\delta e(t,\mathbf{x}) \, , \qquad 
u^\mu(t,\mathbf{x}) \simeq (1,\mathbf{u}(t,\mathbf{x})) \, ,
\end{equation}
where $\delta e$ and $\mathbf{u}$ are the first order disturbance. 
In the linearized conformal limit, the dissipative part is given by the first and 
second order derivative of the fields \cite{Baier:2007ix}
\begin{equation}
\pi^{\mu\nu}=
-2\eta\langle \partial^\mu u^\nu\rangle -2\eta\tau_\pi\langle \partial^\mu
\partial^\nu \ln T\rangle \, \qquad \qquad \mbox{(Static)} \,  ,
\label{staticform}
\end{equation}
where $\langle \, \cdots \, \rangle$ denotes the symmetric and traceless spatial component. 
The second order term can be replaced by the leading order relation,  
making it as a dynamical equation  
\begin{equation}
\pi^{\mu\nu}=
-2\eta\langle \partial^\mu u^\nu \rangle -\tau_\pi\partial_t
\pi^{\mu\nu} \, \qquad \qquad  \mbox{(Dynamic)} \,  .
\label{dynamicform}
\end{equation}
With the constituent relation, we solve the equation of motion in presence of the external force (minus the drag force) 
\begin{equation}
\partial_\mu T^{\mu\nu}=F_{micro}^\nu \, .
\end{equation}
In this case, the force 
$F_{micro}^\nu=\rm{d}p^\nu /\rm{d}t$ is given by
\begin{equation}
F_{micro}^\nu
=\int_{\mathbf{p}}p^\nu S(t,\mathbf{x},\mathbf{p})
=\mu_F v^\nu\delta^3(\mathbf{x}-\mathbf{v} t) \, ,
\end{equation}
where $v^\nu=(v^2,\mathbf{v})$.

At long distances from the quark, the stress-energy tensor is described by 
$T_{hydro}^{\mu\nu}$  
and at short distances, we have to consider correction $\tau^{\mu\nu}$ 
which is localized near the quark
\begin{equation}
T^{\mu\nu}=T_{hydro}^{\mu\nu}+\tau^{\mu\nu}.
\end{equation} 
Then the equation of motion in Fourier space becomes
\begin{equation}
-i\omega \, \delta T^{0j}+ik^i \, \delta T_{hydro}^{ij}
=F_{micro}^j-ik^i\tau^{ij}
\end{equation}
and the correction part acts as a source. 
In the second order, $\tau^{ij}$ can be written with three symmetric tensors 
consisting of $\mathbf{v}$ and $\mathbf{k}$ 
\begin{eqnarray}
\tau^{ij}
&=&2\pi\mu_F \, \delta(\omega-\mathbf{v}\cdot\mathbf{k}) \,
\left[ \, \left(v^iv^j-\frac{1}{3}v^2\delta^{ij}\right) \,
\phi_1(\omega,k^2) \right.\nonumber\\
&& \qquad  \qquad   \qquad  \qquad  \quad
+\left. \left(iv^i k^j+ik^{i}v^{j}-\frac{2}{3}i(\mathbf{v}\cdot\mathbf{k})
\delta^{ij}\right) \, \phi_2(\omega,k^2)
+\left(k^ik^j-\frac{1}{3}k^2\delta^{ij}\right) \, 
\phi_3(\omega,k^2) \, \right] , \qquad 
\end{eqnarray}
where $\phi_i(\omega,k^2)$ $(i=1,2,3)$ is analytic near $\omega=\mathbf{k}=0$. 
Since $\tau^{ij}$ is localized, 
we can expand it for small $\omega$ and $\mathbf{k}$ 
\begin{eqnarray}
-ik^i\tau^{ij}&=&2\pi\mu_F \, \delta(\omega-\mathbf{v}\cdot\mathbf{k})
\left[ \, -i\phi_1(0,0)\, \omega v^j
+\frac{1}{3}iv^2\phi_1(0,0)\, k^j
+\frac{\partial\phi_1(0,0)}{i\partial \omega}\omega^2v^j
\right.\nonumber\\
&& \qquad   \qquad  \qquad   \qquad  \quad
+\left.\frac{1}{3}\left(-v^2\frac{\partial\phi_1(0,0)}{i\partial \omega}
+\phi_2(0,0)\right)\omega k^j+\phi_2(0,0)k^2v^j \, \right]
+\mathcal{O}(k^3) \, .
\end{eqnarray}
We determine $\tau^{\mu\nu}$ by comparing the numerical solution $T^{\mu\nu}$ 
with the constituent relation $T_{hydro}^{\mu\nu}$. 
By comparing order by order, 
all of the coefficients are numerically determined. 
~Table~\ref{sourcecoeff} shows the result where the coefficients are 
measured compared to the shear length unit $[R]=[(4\eta/3)/(c_s^2sT)]$. 
Note that $\phi_1(0,0)$ and $\frac{\partial\phi_1(0,0)}{i\partial\omega}$ 
vanish for the Boltzmann equation. 
Since the source $S(\mathbf{p})$ in Eq.~(\ref{source})
 does not have $H_{l,\pm 2}$ components which 
correspond to the tensor $\left[v^iv^j-\frac{1}{3}v^2\delta^{ij}\right]$, 
there is no $\phi_1(\omega,k^2)$ in the Boltzmann case.

\begin{table*}[t]
\begin{center}
\caption{The correction to the source. 
$[R]=\left[\frac{4\eta/3}{c_s^2sT}\right]$ }
\begin{tabular}{|l||c|c|c|}
\hline & \, $\phi_1(0,0)$  [$R$] \, & 
\, $\frac{\partial\phi_1(0,0)}{i\partial \omega}$  [$R^2$] \, & 
\, $\phi_2(0,0)$  [$R^2$] \,  \\
\hline 
\, Boltzmann \, & 0 & 0 & 0.48 \\ 
\hline 
\, AdS/CFT \, & -1 & 0.34 & -0.33 \\
\hline
\end{tabular}
\label{sourcecoeff}
\end{center}
\end{table*}

\begin{figure}[h]
\centering
\includegraphics[width=80mm]{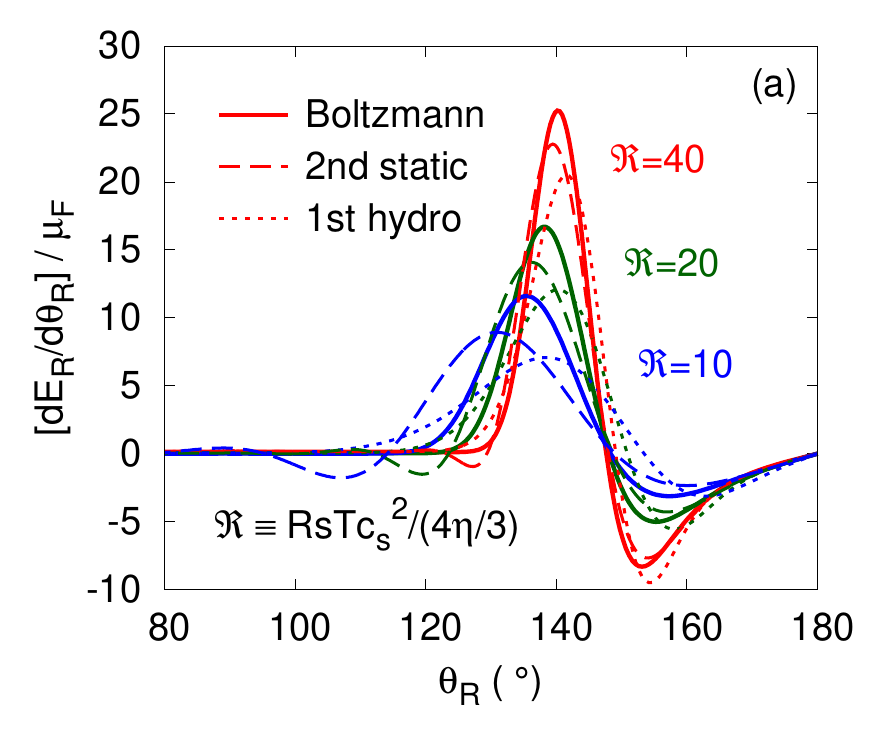}
\includegraphics[width=80mm]{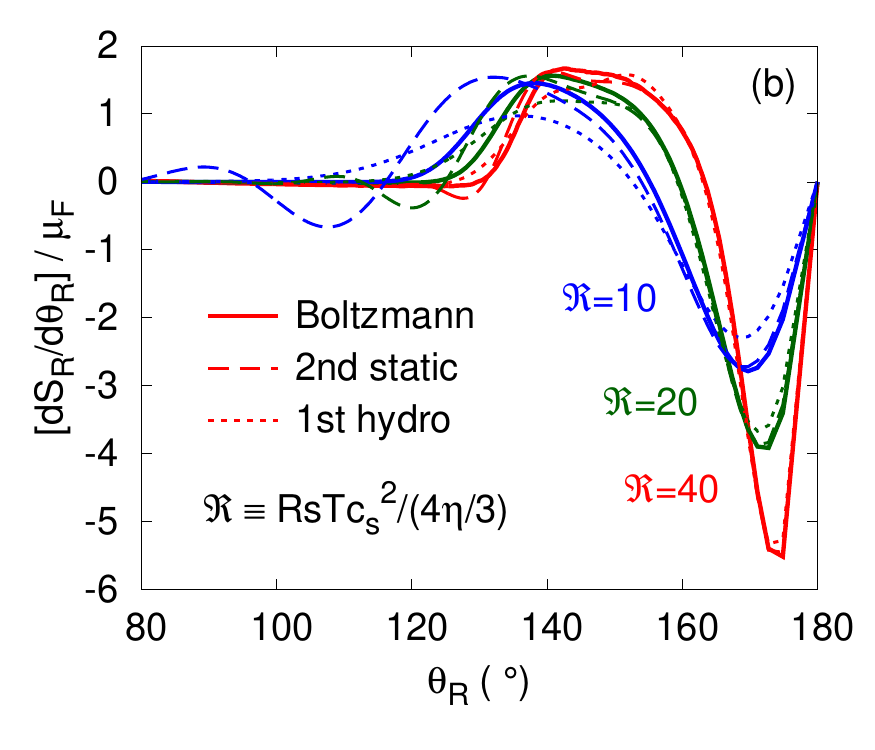}
\caption{The angular distribution of (a) the energy density 
$[\rm{d}E_R/\rm{d}\theta_R]/\mu_F$ and   
(b) the energy flux $[\rm{d}S_R/\rm{d}\theta_R]/\mu_F$
given by the Boltzmann equation at distances 
$\mathcal{R}=10, \, 20, \, \mbox{and} \, 40$. 
The Boltzmann results are compared with the first order and the second order static hydrodynamics.}
\label{Boltzh}
\end{figure}

\begin{figure}[h]
\centering
\includegraphics[width=80mm]{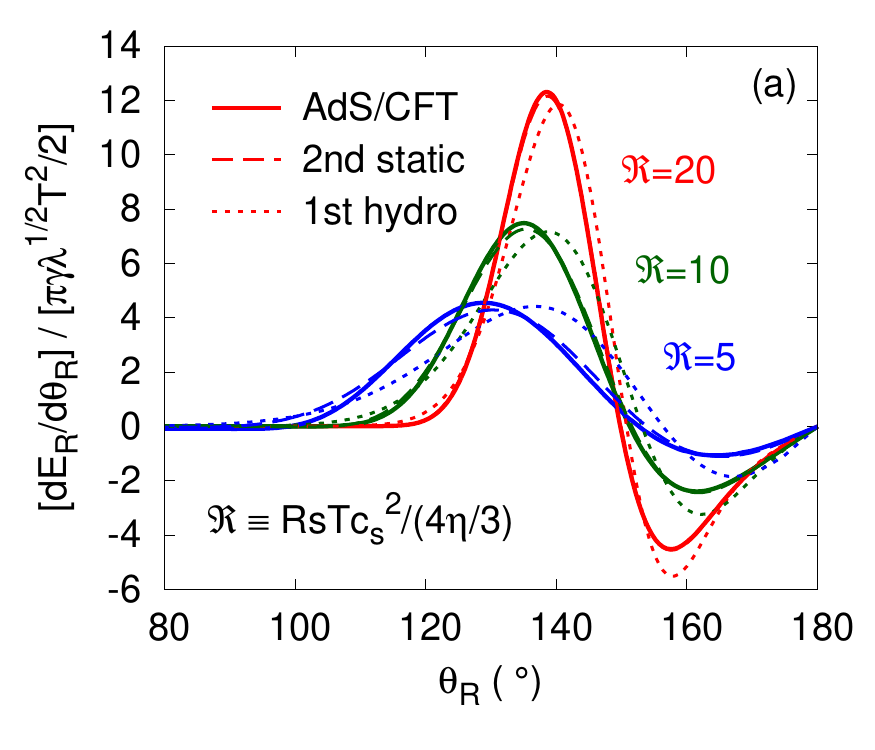}
\includegraphics[width=80mm]{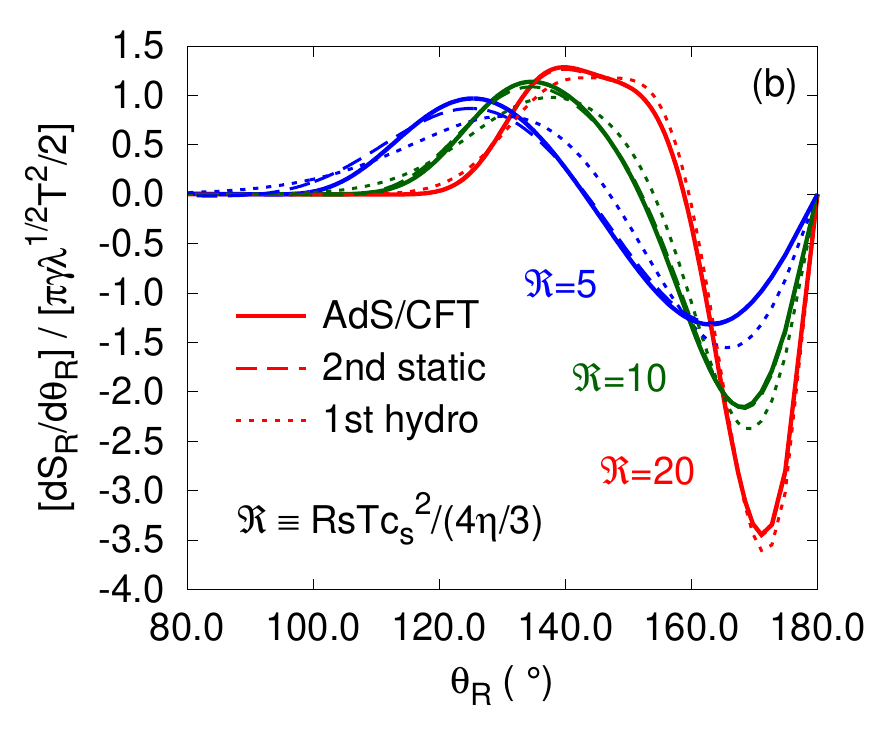}
\caption{The angular distribution of (a) the energy density 
$[\rm{d}E_R/\rm{d}\theta_R]/[\pi \gamma \sqrt{\lambda}T^2/2]$ 
and (b) the energy flux 
$[\rm{d}S_R/\rm{d}\theta_R]/[\pi \gamma \sqrt{\lambda}T^2/2]$
given by the AdS/CFT correspondence at distances 
$\mathcal{R}=5, \, 10, \, \mbox{and} \, 20$. 
The AdS/CFT results are compared with the first order and the second order static hydrodynamics.  
($\gamma = 1/\sqrt{1-v^2}$)} 
\label{AdSh}
\end{figure}

\begin{figure}[h]
\centering
\includegraphics[width=80mm]{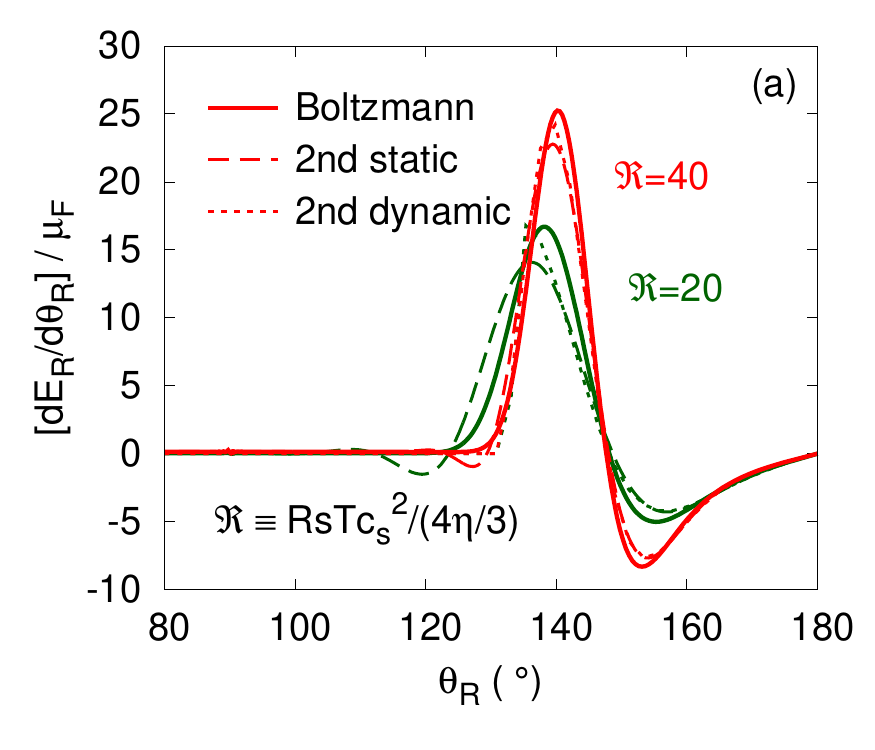}
\includegraphics[width=80mm]{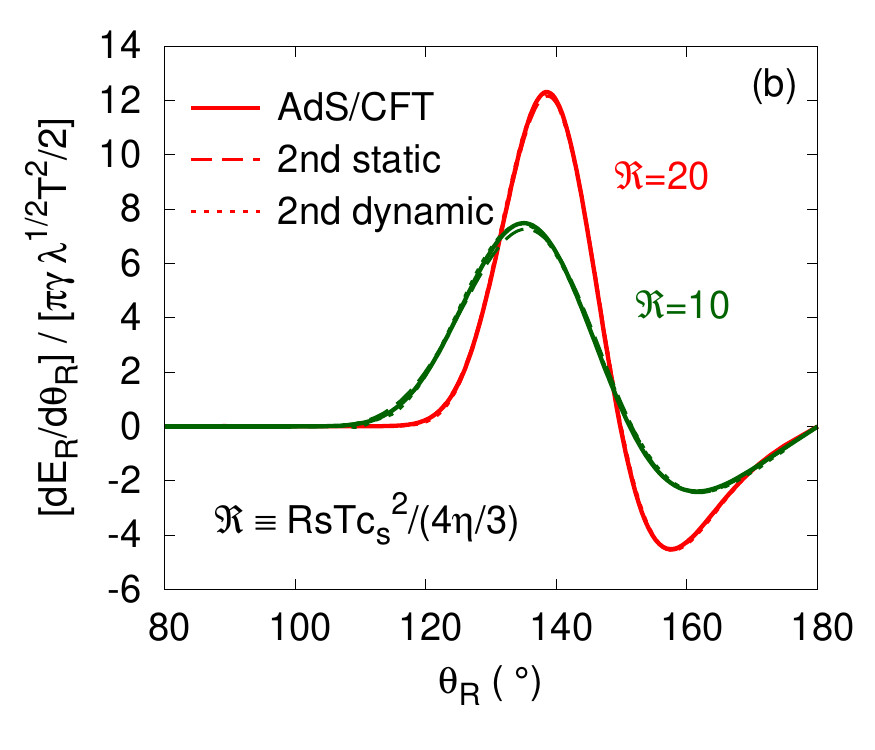}
\caption{(a) The angular distribution of the energy density, 
$[\rm{d}E_R/\rm{d}\theta_R]/\mu_F$ given by the Boltzmann equation 
at distances $\mathcal{R}= 20, \, 40$. 
The Boltzmann results are compared with the second order static and dynamic hydrodynamics.  
The dynamic solution develops spurious shocks for $\mathcal{R}<20$.  
(b) The angular distribution of the energy density, 
$[\rm{d}E_R/\rm{d}\theta_R]/[\pi \gamma \sqrt{\lambda}T^2/2]$ 
given by the AdS/CFT correspondence at distances $\mathcal{R}= 10, \, 20$. 
The AdS/CFT results are compared with the second order static and dynamic hydrodynamics.  
($\gamma = 1/\sqrt{1-v^2}$)} 
\label{2ndhydro}
\end{figure}

With the corrected force 
$F_{micro}^j-ik^i\tau^{ij}\equiv 2\pi\mu_F
\delta(\omega-\mathbf{v}\cdot\mathbf{k})
[\mathbf{v} \, \phi_v(\omega,k^2)+i\mathbf{k}\, \phi_k(\omega,k^2)]$ 
\cite{Chesler:2007sv}, 
we solve the equation of motion with $v=1$  
\begin{eqnarray}
-i\omega \, \delta T^{00}+ik \, \delta T^{0z'}&=&2\pi\mu_F \, 
\delta(\omega-\mathbf{v}\cdot\mathbf{k}) \, ,\nonumber\\
-i\omega \, \delta T^{0z'}+ic^2(k)k \, \delta T^{00}
+\Gamma_s(\omega)k^2 \, \delta T^{0z'}&=&
\left[\cos\theta \, \phi_v+ik \, \phi_k \right]  
2\pi\mu_F \, \delta(\omega-\mathbf{v}\cdot\mathbf{k}) \, , \nonumber\\
-i\omega \, \delta T^{0x'}+D(\omega)k^2 \, \delta T^{0x'}
&=&\sin\theta \, \phi_v \, 2\pi\mu_F \, 
\delta(\omega-\mathbf{v}\cdot\mathbf{k}) \, ,
\end{eqnarray}
where $\phi_v(\omega,k^2)$ and $\phi_k(\omega,k^2)$ are expanded around 
$\omega=\mathbf{k}=0$ depending on the order and  
$c^2(k)=c_s^2=1/3$ is the speed of sound, 
$\Gamma_s(\omega)=\Gamma_s=(4\eta/3)/(sT)$ is the sound attenuation length, 
and $D(\omega)=D=\eta/(sT)$ is the diffusion constant 
in the first order hydrodynamics. 
In the second order static hydrodynamics (Eq.~(\ref{staticform})), 
we have same constants except $c^2(k)=c_s^2(1+\tau_\pi\Gamma_sk^2)$ 
while the exception for the dynamic case (Eq.~(\ref{dynamicform})) 
is $\Gamma_s(\omega)=\Gamma_s/(1-i\tau_\pi\omega)$ and 
$D(\omega)=D/(1-i\tau_\pi\omega)$.  
The hydrodynamic solutions are given by 
\begin{eqnarray}
\delta T^{00}(\omega,\mathbf{k})
&=&\frac{i\left[\omega+k\cos\theta\, \phi_v\right]
-k^2\left[\Gamma_s(\omega)+\phi_k\right]}
{\omega^2-c^2(k)\, k^2+i\Gamma_s(\omega) \, \omega k^2}
\, 2\pi \mu_F \delta(\omega-k\cos\theta) \, , \nonumber\\
\delta T^{0z'}(\omega,\mathbf{k})
&=&\frac{i\left[ \, \omega\cos\theta\, \phi_v 
+c^2(k) \, k \, \right]
-k\omega \, \phi_k}
{\omega^2-c^2(k) \, k^2+i\Gamma_s(\omega) \, \omega k^2}\,
 2\pi \mu_F \delta(\omega-k\cos\theta) \, ,\nonumber\\
\delta T^{0x'}(\omega,\mathbf{k})
&=&\frac{i\sin\theta\, \phi_v}
{\omega+iD(\omega) \, k^2} \, 
2\pi \mu_F\delta(\omega-k\cos\theta) \, .
\end{eqnarray}
In ~Figure~\ref{Boltzh}, \ref{AdSh}, we compare hydrodynamic results to 
the full theory. 
The difference between the static and the dynamic solution is shown in
 ~Figure~\ref{2ndhydro}.

\section{Discussion}

We presented the energy density and flux distribution 
given by kinetic theory based on the Boltzmann equation 
at weak coupling, and compared them with those 
given by the AdS/CFT correspondence at strong coupling. 
Both of them show a Mach cone and diffusion wake at long distances but 
their behaviors are distinguishable by the approach to the hydrodynamic limit. 
 
In the Boltzmann equation, the second order static hydrodynamics 
(Eq.~(\ref{staticform})) works better than 
first order for $\mathcal{R}\gtrsim 40$, 
but it is unnatural for $\mathcal{R}\lesssim 10$ 
due to the high frequency behavior. 
In principle, the second order dynamic solution (Eq.~(\ref{dynamicform}))
 is better than static one at long distances
(see ~Figure~\ref{2ndhydro} (a)). 
However, the dynamic solution has higher order terms and 
develops spurious shocks for $\mathcal{R}<20$. 

In the AdS/CFT correspondence, 
even the first order hydrodynamics is good for $\mathcal{R}\gtrsim 5$. 
Only the sound wave has corrections of the second order static hydrodynamics  
and is dominant in the energy distribution. 
At $\mathcal{R}=5$ in the energy density, 
we can see the improvement given by the second order theory. 
Both the static and the dynamic solution work very well 
in ~Figure~\ref{2ndhydro} (b). 

In comparison with the AdS/CFT correspondence, 
the Boltzmann equation produces a smooth transition 
from the free streaming to the hydrodynamic regime. 
The stress-energy tensor given by the AdS/CFT correspondence 
is significantly better described by hydrodynamics 
in the sub-asymptotic regime. 
Although they have different sources, we believe that 
this difference is due to the second order hydrodynamic 
coefficient, $\tau_\pi$. 
Compared to the shear length, they are given by
\begin{equation}
\tau_\pi \simeq  1.58 \left[\frac{4\eta/3}{c_s^2sT}\right] 
\quad \mbox{(Boltzmann)} \, \quad \qquad  \longleftrightarrow  \qquad \quad
\tau_\pi \simeq  0.65 \left[\frac{4\eta/3}{c_s^2sT}\right]
\quad \mbox{(AdS/CFT)} \, .
\end{equation}
The coefficient $\tau_\pi$ in unit of $\eta/(sT)$ is 
generically $\sim2.4$ times larger in the Boltzmann equation 
than in gauge/gravity duality. 
As a result, the hydrodynamic solutions are more reactive 
to the high frequencies and it takes longer to reach the hydrodynamic regime, 
in a relative sense, in the Boltzmann equation.

\begin{acknowledgments}
This work is done in collaboration with Derek Teaney and Paul M. Chesler and 
supported by an OJI grant from the Department of Energy 
DE-FG-02-08ER4154 and the Sloan Foundation. 

\end{acknowledgments}

\bigskip 

\begin{thebibliography}{9}   


\bibitem{Gubser:2007ga}
S.~S.~Gubser and S.~S.~Pufu and A.~Yarom, 
Phys.\ Rev.\ Lett. {\bf 100}, 012301 (2008)
[arXiv:0706.4307 [hep-th]]. 

\bibitem{Chesler:2007sv}
P.~M.~Chesler and L.~Yaffe,
Phys.\ Rev.\ D {\bf 78}, 045013 (2008)
[arXiv:0712.0050 [hep-th]].

\bibitem{Arnold:2000dr}
P.~Arnold, G.~D.~Moore and L.~G.~Yaffe,
JHEP {\bf 0011}, 001 (2000)
[arXiv:hep-ph/0010177].

\bibitem{Hong:2010at}
J.~Hong and D.~Teaney,
Phys.\ Rev.\ C {\bf 82}, 044908 (2010)
[arXiv:1003.0699 [nucl-th]].

\bibitem{Arnold:2006fz}
P.~Arnold, C.~Dogan and G.~D.~Moore,
Phys.\ Rev.\  D {\bf 74}, 085021 (2006)
[arXiv:hep-ph/0608012].

\bibitem{Kovtun:2004de}
P.~Kovtun, D.~T.~Son and A.~O.~Starinets,
Phys.\ Rev.\ Lett. {\bf 94}, 111601 (2005)
[arXiv:hep-th/0405231].

\bibitem{Baier:2007ix}
R.~Baier, P.~Romatschke, D.~T.~Son, A.~O.~Starinets, and M.~A.~Stephanov, 
JHEP {\bf 04}, 100 (2008)
[arXiv:0712.2451 [hep-th]].

\end{thebibliography}

\end{document}